\documentclass[twocolumn,showpacs,preprintnumbers,amsmath,amssymb,prb]{revtex4}
\usepackage{graphicx}
\usepackage{dcolumn}
\usepackage{bm}

\newcommand{\D}{\mbox{$D$}}

\def\etal{{\it et al}.}

\begin{document}
\normalsize
\begin{minipage}[10 in]{6.75 in}
\large
\noindent
\textbf{emiT: an apparatus to test time reversal invariance in polarized neutron decay}\\
\normalsize
\\\\
H.\,P.~Mumm, A. Garcia, L.~Grout, M.~Howe, L.\,P.~Parazzoli, R.\,G.\,H.~Robertson, K.\,M.~Sundqvist, and J.\,F.~Wilkerson\\
{\it CENPA, University of Washington, Seattle, WA 98195}
\vspace{.25 in}\\
S.~J.~Freedman, B.~K.~Fujikawa, and L.~J.~Lising~\cite{LJLADD}\\
{\it Physics Department, University of California at Berkeley and Lawrence Berkeley 
National Laboratory, Berkeley, CA 94720}
\vspace{.25 in}\\
M.~S.~Dewey, J.~S.~Nico, and A.~K.~Thompson\\
{\it National Institute of Standards and 
Technology, Gaithersburg, MD, 20899}
\vspace{.25 in}\\
T.~E.~Chupp, R.~L.~Cooper, K.~P.~Coulter, S.~R.~Hwang, and R.~C.~Welsh\\
{\it Physics Department, University of Michigan, Ann Arbor, MI, 48104}
\vspace{.25 in}\\
L.~J.~Broussard, C.~A.~Trull and F.~E.~Wietfeldt\\
{\it Physics Department, Tulane University, New Orleans, LA 70118}
\vspace{.25 in}\\
G.~L.~Jones\\
{\it Physics Department, Hamilton College, Clinton, NY 13323}\\
\vspace{4.2 in}
\end{minipage}
\renewcommand{\baselinestretch}{1.0}
\normalsize
\begin{abstract} 

We describe an apparatus used to measure the triple-correlation term (\D\,$\hat{\sigma}_{n}\cdot \bf{{p_e}}\times\bf{{p_\nu}}$) in the beta-decay of polarized neutrons.  The \D-coefficient is sensitive to  possible violations of time reversal invariance.  The detector has an octagonal symmetry that optimizes electron-proton coincidence rates and reduces systematic effects.  A beam of longitudinally polarized cold neutrons passes through the detector chamber, where a small fraction beta-decay.  The final-state protons are accelerated and focused onto arrays of cooled semiconductor diodes, while the coincident electrons are detected using panels of plastic scintillator.  Details regarding the design and performance of the proton detectors, beta detectors and the electronics used in the data collection system are presented.  The neutron beam characteristics, the spin-transport magnetic fields, and polarization measurements are also described.

\end{abstract}

\maketitle

\section{Introduction}\label{intro}

Charge-parity symmetry violation (CP violation) is an important property of nature. Of particular interest is that it is needed to explain the preponderance of matter over antimatter in the universe~\cite{SAK91}. Thus far, CP violation has been observed only in K and B meson oscillation and decays~\cite{CHR64,AUB01,ABE01} and  can be entirely accounted for by a phase in the Cabbibo-Kobayashi-Maskawa (CKM) matrix in the electroweak Lagrangian. It can be shown, however, that this phase is insufficient to account for the known baryon asymmetry in the context of Big Bang cosmology~\cite{RIO99} so there is good reason to search for CP violation in other systems.  CP and Time-reversal (T) violation can be related to each other through the Charge-Parity-Time (CPT) theorem.  Experimental limits on neutron and atomic electric dipole moments (T-violating) place strict constraints on some, but not all, possible sources of new CP violation.  Tests of nuclear beta decay, and neutron decay in particular, complement these experiments. Some theoretical models that extend the Standard Model, such as left-right symmetric theories, leptoquarks, and certain exotic fermions could cause observable effects that are as large as the present experimental limits~\cite{HER98}.
\par
The decay probability distribution for neutron beta decay $dW$, written in terms of 
the neutron spin direction $\hat{\sigma}_{n}$ and the momenta (energy) of the electron $\bf{p_e}$ ($E_e$) and antineutrino $\bf{p_{\nu}}$ ($E_\nu$) was first described by Jackson, Trieman, and 
Wyld in 1957~\cite{JAC57}, 

\begin{equation}
\begin{array}{lr}
\label{jtweqn}
dW \propto \bigg(&1+ a{{\bf{p_e}}\cdot{\bf{p_\nu}}\over {E_eE_\nu}}
+ A{\hat{\sigma}_{n}\cdot{\bf{p_e}} \over {E_e}} + B{\hat{\sigma}_{n}
\cdot{\bf{p_\nu}} \over {E_\nu}} \\
& + D{\hat{\sigma}_{n}\cdot {\bf{p_e}}\times{\bf{p_\nu}}\over {E_eE_\nu}}
\  \bigg)dE_ed\Omega_{e}d\Omega_{\nu}.
\end{array}
\end{equation}

\begin{figure*}[t]
\includegraphics[width=5.5in]{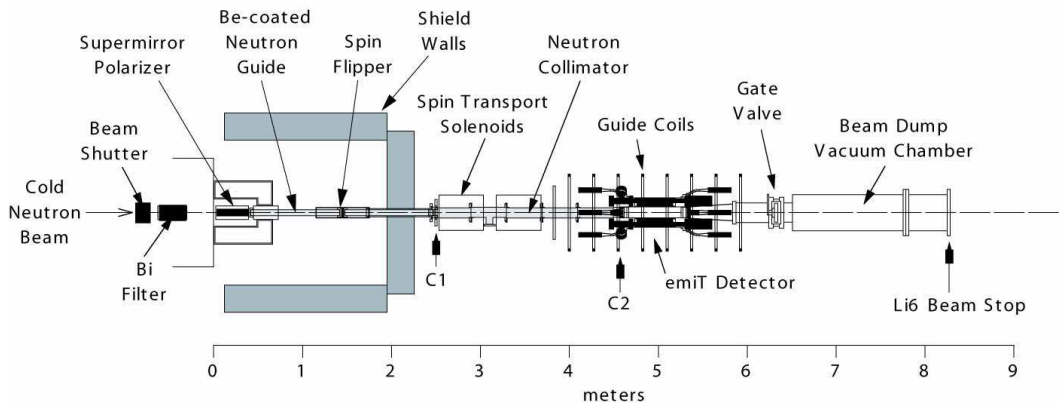}
\caption{The emiT beam line on NG6.}
\label{fig:beamline}
\end{figure*}

The triple correlation $D$ is T-odd.  A non-zero value for \D\ above the level of calculable final state interactions implies a violation of time-reversal symmetry.  The most sensitive measurement of \D\  in $^{19}$Ne decay is $(1\pm6)\times10^{-4}$~\cite{CAL85}, only about a factor of three above electromagnetic final state effects.  The most recent measurements of \D\ in neutron decay, $(-2.8 \pm 7.1) \times 10^{-4}$ and $(-6 \pm 13) \times 10^{-4}$ come from the TRINE collaboration and the first run of the emiT experiment respectively~\cite{SOL04, SOL01, LJL99}. Standard Model final state effects in the neutron system are estimated to be approximately $1.3\times10^{-5}$, well below the sensitivity of current experiments~\cite{CAL67}.    Details of the first emiT measurement may be found elsewhere~\cite{LIS99,HWA98}.  The focus of this paper is the description of the upgraded emiT apparatus.  We describe the cold neutron beam line, the spin transport, and specific aspects of an extensively upgraded detector used to perform a second, more sensitive measurement, from October 2002 through January 2004.   

\section{Cold Neutron Beam}\label{nbeam}

The emiT experiment uses polarized cold neutrons at the National Institute of Standards and Technology Center for Neutron Research (NCNR).  The NCNR operates a 20 MW research reactor that provides a source of fission neutrons that have been moderated to thermal energies by the D$_{2}$O primary reactor coolant.  Cold neutrons are produced by a neutron moderator situated adjacent to the reactor core. The cold source was recently upgraded and consists of an ellipsoidal shell of liquid hydrogen maintained at a temperature of 20~K~\cite{WIL03}. Scattering of neutrons in the cold source cools  the neutrons to approximately 40~K (they do not reach thermal equilibrium with the liquid hydrogen).  Colder neutrons spend a longer time within the sensitive volume of the detector, increasing the probability of an observable decay.

Neutron guides coated with $^{58}$Ni efficiently transport the cold neutrons approximately 68~m from the cold source to the experimental area at the end of neutron guide~6 (NG-6)~\cite{HUF03} on the NCNR Neutron Guide Hall floor.  The highly collimated cold neutron beam exits the guide through a thin Mg window and travels through 79~cm of air to an aperture that reduces the beam to a 6~cm diameter cylinder. Downstream of this aperture is the remote-controlled local beam shutter (see Figure~\ref{fig:beamline}).  Upon exiting the guide shutter, the neutron beam passes through a meter-long air gap before entering a 15~cm thick cryogenically cooled beam filter constructed from blocks of single-crystal bismuth.  The filter attenuates fast neutrons and gamma rays originating from the reactor core that would otherwise contribute to the background. Cooling the filter elements to liquid-nitrogen temperatures significantly increases the transmission of cold neutrons through the filter by reducing losses from phonon scattering.  The neutrons exit the end of NG6, pass through a polarizer (Section \ref{npol}), and travel one meter to the spin-flipper (Section \ref{spintrans}) through a Be-coated glass neutron guide tube in which a slight helium overpressure is maintained to prevent beam attenuation due to air scattering. The windows on each end of this guide tube are 0.5~mm thick single-crystal Si.  While in the spin-flipper, the beam passes through two parallel sheets of 0.5~mm  Formvar-coated aluminum wire. The main vacuum chamber begins just past the spin-flipper, with a second 1~m long Be-coated glass neutron guide. The upstream vacuum window of this guide is 0.10~mm aluminum. Following the second guide is the beam collimator section. Two $^6$LiF collimators, C1 (6.00~cm diameter) and C2 (5.00~cm diameter) in Figure~\ref{fig:beamline}, are separated by 2 meters, and define the beam. Between C1 and C2 are four $^6$LiF beam ``scrapers'' with decreasing diameters from 5.90~cm to 5.32~cm. Backing each collimator and scraper is a thick ring of high-purity lead which removes gamma rays and fast neutrons. Between the scrapers, the collimator tube is lined with  $^6$Li-loaded glass to absorb scattered neutrons.  Beyond C2, the beam enters the 80~cm long detector chamber, where the neutron decay products are observed.  Downstream of the detector chamber, the beam travels 2.8 m through vacuum to the $^6$Li-loaded-glass beam stop and fluence-monitoring fission chamber.

\subsection{Neutron Intensity Distribution}\label{nintensity}

Knowledge of the intensity distribution of the neutron beam is essential for understanding potential systematic effects that may give rise to a false \D-coefficient.  We employed a neutron imaging technique to profile the beam at three locations along the beam line, thus obtaining detailed information on the beam envelope.  In this method, the neutron beam irradiates a metal foil with the requirements of a high thermal neutron absorption cross-section,  a decay branch into beta particles,  few competing decay modes, and  a convenient half-life.  Although there are a number of suitable metals for use as the transfer foil, we used natural dysprosium, the relevant isotope being $^{164}$Dy.  After irradiation, the decay electrons from the activated foil expose a film that is sensitive to beta particles and can be read out by an image reader.  Typically  the pixel resolution of the images obtained was $200~\mu$m by $200~\mu$m.  The intrinsic resolution of the image can be better but is not needed for this application.  This technique has not been commonly used for neutron imaging, however further examples may be found in Refs.~\cite{CHE96,CHE00,CHO00}.

\begin{figure}
\includegraphics[width=3.in]{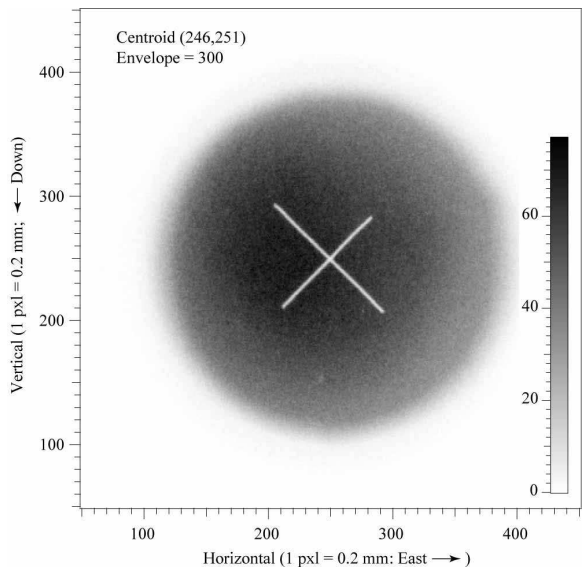}
\caption{An image plot of the neutron beam intensity profile obtained from a dysprosium foil.
The ``$\times$'' in the center of the beam indicates the mechanical axis of the detector and is centered at position (250,250). The centroid of the image occurs at position (246,251).}
\label{fig:beamimage}
\end{figure}

Beam images were obtained at the axial center of the emiT detector as well as $18$~cm  upstream and downstream. These images allow a  rendering of the beam envelope sufficient for our purposes. Figure~\ref{fig:beamimage} shows an example of one of the images. The intensity scale, referred to as photostimulable luminescence, is linearly proportional to the neutron fluence.  The film is initially read-out on a logarithmic scale, which covers about four decades of dynamic range (making it ideal for sensitive neutron measurements).  One converts to a linear scale through a function supplied by the manufacturer of the film reader.  The ``$\times$'' in the center of the beam indicates the mechanical axis of the detector. It was obtained by sighting with a theodolyte and mounting thin Cd wire on the dysprosium; it is centered at position (250,250). The centroid of the image was obtained by weighting each position by its intensity. The centroid defines the beam axis.  For all three images, the centroid occurs within $\pm1$~mm of the mechanical center (detector axis).  Figure~\ref{fig:beamimage}  indicates that the peak of the beam is not coincident with the centroid. This asymmetry arises from the reflection of neutrons from the polarizing supermirror (PSM) discussed in Section~\ref{npol}.  It is possible to force the centroid and peak to coincide by rotating the PSM; unfortunately the beam profile remains asymmetric, and there is an undesirable 25\% loss in polarized neutron fluence (neutrons per square centimeter).  The maximum neutron fluence corresponds to matching of the phase space of the beam with PSM acceptance.  Since the PSM is curved, this places the beam axis at an angle to the PSM so that the beam divergence is asymmetric. 

Because the precise wavelength distribution of the polychromatic beam is unknown, one uses the capture fluence to quantify the neutron density $\rho v_o$, where $v_o = 2200$~m/s.  Using a calibrated fission chamber, the capture fluence rate was measured to be  $1.4 \times 10^9$ cm$^{-2}$s$^{-1}$ at the entrance to the polarizer and  $1.7 \times 10^8$ cm$^{-2}$s$^{-1}$ at C2. 

\subsection{The Neutron Polarization}\label{npol}
\label{npolar}
Spin-polarization of the neutron beam is achieved using a PSM~\cite{SCH89,SC89} obtained from the Institut Laue-Langevin (ILL) in Grenoble, France.  Supermirrors produce a high degree of polarization, typically greater than 95\%, and are quite stable. The PSM consists of forty 0.2~mm $\times$ 6~cm $\times$ 33~cm Pyrex{\scriptsize$^{\textregistered}$}~plates with coatings on each side that maximize the reflection of the desired spin state while absorbing nearly all of the other~\cite{PYR03, Disclaimer}.  Both sides of each plate are covered with several reflecting layers of cobalt and titanium, each layer a few angstroms thick. Beneath those are layers of gadolinium and titanium, which further refract and absorb the neutrons of the undesired spin state.  Each plate is slightly curved with a radius of 10~m so that there is no line of sight through the polarizer; each neutron transmitted reflects at least once.

The PSM is positioned within the gap of a magnet assembly formed by two rows of permanent magnets that are rigidly mounted with steel plates.  This entire assembly is kinematically mounted on precision translational and rotational stages, allowing it to be precisely aligned relative to the neutron beam line.  The permanent magnets produce a field within the gap of approximately 20~mT, transverse to the direction of neutron flight.  Neutrons of the desired spin state are reflected via the spin-dependent scattering cross-section from the magnetic layer, while neutrons of the opposite spin state are captured on the gadolinium.  In this manner, the emergent neutron beam is spin-polarized to better than 90\%.  The polarizer has an overall transmission of 24\% of the neutron fluence incident on the PSM.  Because neutron capture in gadolinium results in the emission of a large number of capture gamma rays, the PSM polarizer is shielded with 18~cm of lead in order to minimize the gamma-ray background present at the emiT detector. 

\subsubsection{Polarimetery}
The neutron beam polarization was measured during construction of the emiT beam line.  An analyzing supermirror (ASM), similar in design to the PSM but with supermirror coatings on only one side of each glass plate, was set up at the position of the second collimator C2. A fission chamber was mounted on the downstream side of the analyzer.  The product of the beam polarization, $P$, and the analyzing power of the ASM, $A$, is given by
\begin{equation}
AP={N_{u}-N_{f}\over{sN_{u}+N_{f}}},
\end{equation}
\noindent where $N_{u}$ is the number of counts obtained with the beam polarization in one spin state (``no-flip''), and $N_{f}$ is the number of counts obtained with the beam polarization in the opposite spin state (``flip'').  The spin-flip efficiency $s$ is the absolute value of the ratio of polarizations for the two states of the spin-flipper. This was not measured but our calculations indicate $s=0.95 \pm 0.05$~\cite{HWA98}.  Single-sided bender polarizers are known to have a smaller polarizing power than the double-sided~\cite{SC89} so we can reasonably assume that $A < P$.  We measured
\begin{equation}
\frac{N_{u}}{N_{f}} = 10.820 \pm 0.02.
\end{equation}
Using the extreme values of $s = 1$ and $A = P$, we obtain a lower limit on the beam
polarization of $P > 0.91$~\cite{FLIP}.  Incomplete polarization simply reduces the sensitivity of the measurement, and uncertainty in the polarization leads to a measurement uncertainty proportional to \D.  Precise polarimetery is therefore unnecessary.  

\subsubsection{Spin Transport}\label{spintrans}

The magnetic fields downstream of the supermirror are designed to allow the neutron polarization to be flipped, to maintain the neutron polarization into the detector region, and to provide a precisely aligned quantization axis in the detector region.  Exiting the PSM, the neutrons proceed through a double current-sheet spin-flipper.  The current sheets are 30~cm $\times$ 12~cm (vertical $\times$ horizontal) with horizontal currents and are wound with close packed 0.45~mm Al wire. The return wires for each current sheet are 30~cm upsteam/downstream with the central 10 cm of windings bent vertically to allow space for the guide tubes and spin transport solenoids. Iron plates are used to maintain the vertical fields between the supermirror and the upstream current-sheet.

The current in the downstream current-sheet can be in the opposite direction as the upstream one so that the magnetic field at the 1~mm transition is largely determined by the return coils of each current sheet (no-flip state).  Alternatively its current can be in the same direction, causing a sharp reversal in the field direction between the two current-sheets (flip state).  In the no-flip state, the fields are aligned and the neutrons see no change of field going through the spin flipper. In the flip state, the neutrons see an abrupt transition from 2.5~mT  to -2.5~mT when they cross the gap between the solenoids (there is a maximum residual field of 0.2~mT at the edge of the neutron beam).  A 0.4~nm neutron traverses the field flip in approximately 1~$\mu$s.  Thus the field rotation exceeds the Larmor precession rate, and the direction of the field is reversed but the spin direction is not.  In this way the polarization along the magnetic field is reversed. The spin flip efficiency is estimated to be 95\%  from numerical integration of the Bloch equations describing the evolution of free spins in a magnetic field. (This integration also shows that any residual transverse magnetization left from an incomplete spin flip quickly averages to zero over the neutron decay region because of the polychromatic beam.)

\begin{figure}
\includegraphics[width=3.in]{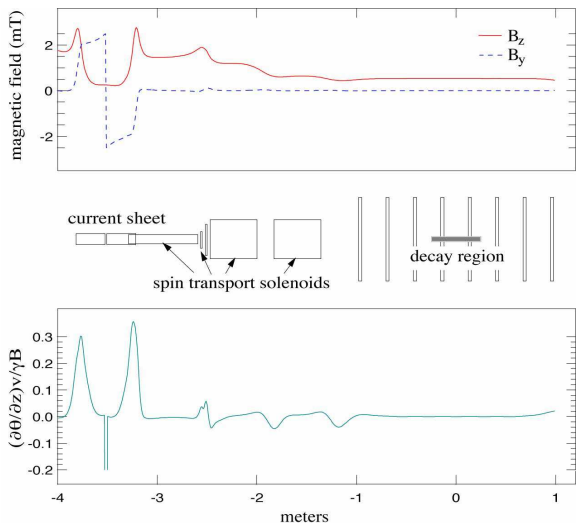}
\caption{Calculated spin transport fields.  The top two traces show the $y$ (vertical) and $z$ (along the beam line) components of the spin transport field respectively.  The bottom trace shows the adiabatic parameter $(\partial\theta/\partial z)v/\gamma B$.  At the center of the figure is a layout of the spin transport components.}
\label{fig:adiabat}
\end{figure}

Once the neutrons exit the downstream flipper solenoid, their spins are adiabatically rotated into the axis of a solenoid 70~cm long by 9.5~cm in diameter that is concentric with the beam line. The neutrons travel through three loops and two more axially aligned solenoids (50~cm $\times$ 41~cm diameter) before entering the detector region.  The 41~cm solenoids provide a smooth transition into the primary field in the detector region that is formed by eight 0.82~m diameter coils equally spaced along 2~m of the beam line.  The guide field in the detector region is approximately 0.5~mT.  Figure~\ref{fig:adiabat} shows the guide fields as a function of position along the beam line.  Also shown is the value of the adiabaticity parameter comparing the gradient of the magnetic field to the Larmor precession rate.  

\subsubsection{Field Alignment}
Precise alignment of the beam axis, the detector axis, and the magnetic field is critical to minimizing systematic effects.  A misalignment between the detector axis and the neutron spin axis can mimic a time-reversal violating signal if the beam is not centered in the detector. This false signal is due to a combination of the beta and neutrino asymmetries ($A$ and $B$), solid angle, and decay kinematics. To reduce systematic effects to acceptable levels, both the detector and the field are aligned to the beam line to within a few miliradians. The detector could be physically aligned using cross-hairs temporarily mounted in the ends of the decay chamber, but the field required a more complicated alignment procedure described below. 

The beam axis is established as described in Section \ref{nintensity}. The alignment of  the magnetic field is accomplished using a Bartington $Mag$-03MS1000 three axis fluxgate magnetometer.  As the internal alignment of this magnetometer is specified to only 1.8~mrad, an alternative method of alignment was developed. Using an alignment tube with crosshairs, a V-block is aligned to within 1~mrad over the length of a 0.6~m rail mounted in the decay region.  The magnetometer is mounted in a square carriage that could be placed in the aligned V-block. The square sides of the carriage allowed the fluxgate to be mounted in four orientations representing precise 90 degree rotations around the V-block axis. By comparing the transverse field measured in two positions 180 degrees apart, it is possible to determine the actual transverse field.

By reversing the current in an individual coil, as well as rotating the magnetometer about the V-block axis, it is possible to determine the misalignment of each coil and align the field in the center of each coil with respect to the V-block axis to within 1~mrad.  However, we also discovered that the fields from each coil are distorted by steel pieces in a trench that crosses the floor approximately 1~m below the middle of the detector. Even with the individual coils aligned, the magnetization induced in this steel causes a marked field misalignment in the decay region. Since this misalignment depends on the field strength in the guide field coils, all further corrections to the magnetic field are done with the 0.5~mT field in the detector region to keep the induced field constant. 

Once each guide field coil is aligned, there remain transverse fields associated with ambient guide hall fields and with the  induced fields described above. Uniform transverse fields are canceled using six rectangular coils placed in a beam-centered array around the detector. Each coil consists of 208~cm long straight sections 90~cm off axis with split semi-circles of 90~cm diameter on either end.  The semi-circular sections allow space for the detector assembly.  By independently varying the currents in these coils a uniform transverse magnetic field can be established at any azimuthal angle.   The residual transverse field components that can not be compensated by these coils are broken down into polynomial moments, and individual correction coils were wound for the largest of these moments:  $dB_x/dz$, $dB_y/dz$, $d^2B_y/dz^2$, and $dB_y/dy$ where y is vertical, x is horizontal, and z is along the beam line.  Though the actual fields from these correction coils are mostly orthogonal, an iterative procedure is used to produce the best cancelation possible with all the coils and the guide field coils turned on.  The resulting field is measured to be aligned within a few mrad throughout the detector region.  Unfortunately, it is not possible to check the alignment of the fields with the detector in place due to the difficulty of aligning the V-block inside the detector.

\subsubsection{Field Monitoring}\label{fieldmon}

To insure stability in the magnetic systems, the current through each of the coils is monitored several milliseconds after every spin flip.  Two fluxgate magnetometers in fixed positions near the detector are also monitored every 100 seconds.  This data, along with other detector parameters, is fed directly into the data-stream.  In addition, five unaligned V-blocks designed to reproducibly position a fluxgate magnetometer are fixed to the detector frame in several places between 30~cm and 1~m from the detector region.  These V-blocks allow for additional checks of the magnetic field. 

\section{The emiT Detector}\label{detector}

\begin{figure*}[t]
\includegraphics[width=5.5in]{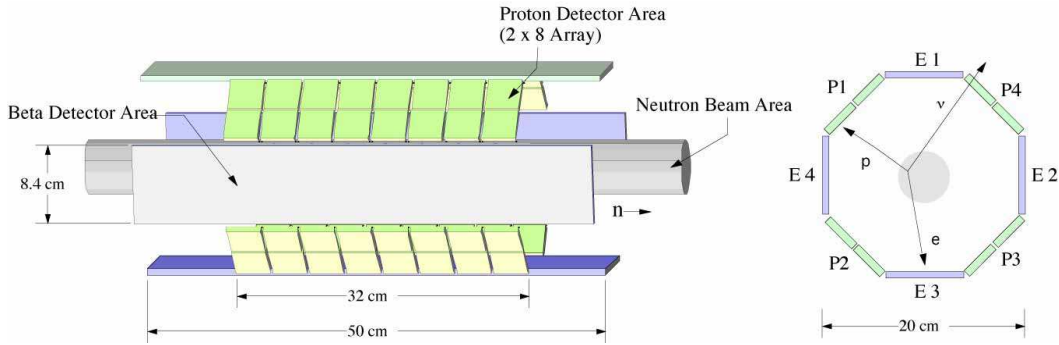}
\caption{A schematic of the emiT detector illustrating the alternating electron and proton detector segments.}
\label{detconfig}
\end{figure*}

The emiT detector consists of an octagonal array of four electron and four proton detectors concentric to a beam of longitudinally polarized neutrons.  The detector configuration is shown schematically in Figure~\ref{detconfig}.   Monte Carlo evaluation of various detector arrangements led to a detector geometry that is simultaneously optimized for sensitivity to the \D-coefficient and insensitivity to most systematic effects, as discussed in following sections~\cite{WAS94}.  The highly symmetric arrangement allows for the approximate cancelation of systematic effects steming from detector efficiency and solid angle variations as well as from beam and polarization misalignment.

\subsection{Detector Design}\label{detdesign}

As discussed in Section~\ref{intro}, the T-violating term is expressed in terms of the neutron spin, electron momentum, and proton momentum as $\hat{\sigma}_{n}\cdot \bf{{p_e}}\times\bf{{p_p}}$.   Short of tracking the individual particles, one can place beta detectors and proton detectors about a beam of polarized neutrons and measure coincidence events from decays within the polarized neutron beam.  One can define left-handed or right-handed events based on the sign  of the triple product, $\hat{\sigma}_{n}\cdot \bf{{p_e}}\times\bf{{p_p}}$.  Although the particle detectors subtend finite solid angles, they can be placed such that nearly all of the neutron decays are correctly identified as either right-handed or left-handed. Monte Carlo studies indicate that while the sensitivity to \D\ relative to other terms in the decay distribution is dependent upon detector geometry, the primary concern should be arranging the coincidence pairs such that the coincidence rate is maximized.

\begin{figure}[h]
\includegraphics[width=3.25in]{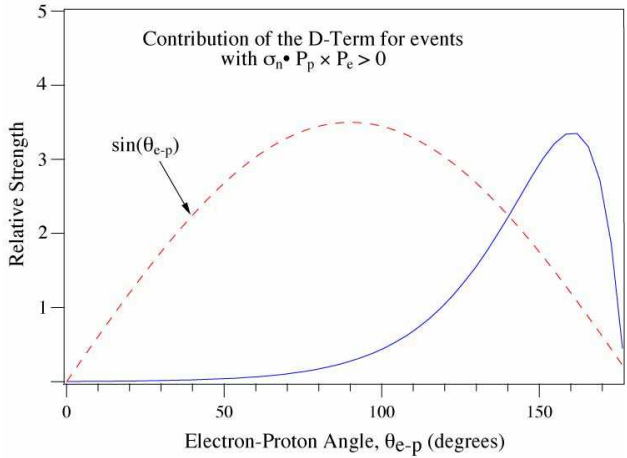}
\caption{Monte Carlo results for the contribution of the \D-coefficient as a function of the beta-proton angle $\theta_{e-p}$ for right-handed events in zero-radius beam.  The solid curve is a fit to the Monte Carlo data.  The dashed curve shows the sine of $\theta_{e-p}$, which would be the contribution to the \D-coefficient if it had only the sine dependence of the cross product.}
\label{decaydistr} 
\end{figure}

The octagonal geometry used in emiT places pairs of beta and proton detectors at an average angle of 135$^{\circ}$ rather than at 90$^{\circ}$ as was typical in earlier experiments.  This choice of angles increases the coincidence detection efficiency because with little decay energy available the electron preferentially recoils from the proton.  The momenta are thus highly anticorrelated and the coincidence rate increases greatly as the angle between the proton and electron approaches 180$^{\circ}$.  This situation is clearly illustrated in Figure~\ref{decaydistr}, which shows a Monte Carlo simulation of the contribution of the triple correlation from Equation~\ref{jtweqn}.  A detector with coincidence pairs subtending the same solid angles but arranged 135$^\circ$ has a greater coincidence count rate than a right angle geometry.  The difference in rate is dependent on the beam diameter (the data shown assumes a zero-diameter beam); for emiT it is  approximately a factor of three. 

\subsubsection{Determining \D}

Since several systematic effects can give rise to a false value of \D, it is useful to outline briefly how one measures the \D-coefficient.  More detailed discussions are found elsewhere~\cite{WAS94,ERO70}.  Note that the analysis presented here assumes that the detector segments have uniform detection efficiency.  This was {\it not} the situation during the run in 1997~\cite{JON99}, and thus, a new analysis method was developed to analyze data where the detector has a high degree of nonuniformity~\cite{LIS99,HWA98}.  The problems that led to this situation have largely been resolved, and the following discussion illustrates a possible analysis approach.

A majority of the possible systematic effects can be eliminated by utilizing a detector with the appropriate symmetry and by reversing the direction of the neutron polarization during the measurement.  Consequently, the emiT experiment is performed by periodically reversing the polarity of the neutron spin, and then comparing the number of coincidences for events with opposite signs of the triple correlation in otherwise identical coincidence pairs of proton and electron detectors.  Using this approach, one can construct a ratio of the number of coincidence events in which all the factors in the numerator favor events with one sign of the triple correlation, and those in the denominator with the opposite sign.  For example, the factors in such a ratio for the geometry shown in Figure~\ref{detconfig} are

\begin{equation}
R={N_{e{_{1}p_{3}}}^{\uparrow}N_{e{_{2}p_{4}}}^{\uparrow}N_{e{_{3}p_{1}}}^{\uparrow}N_{e{_{4}p_{2}}}^{\uparrow}
N_{e{_{1}p_{2}}}^{\downarrow}N_{e{_{2}p_{3}}}^{\downarrow}N_{e{_{3}p_{4}}}^{\downarrow}N_{e{_{4}p_{1}}}^{\downarrow}
\over{N_{e{_{1}p_{2}}}^{\uparrow}N_{e{_{2}p_{3}}}^{\uparrow}N_{e{_{3}p_{4}}}^{\uparrow}N_{e{_{4}p_{1}}}^{\uparrow}
N_{e{_{1}p_{3}}}^{\downarrow}N_{e{_{2}p_{4}}}^{\downarrow}N_{e{_{3}p_{1}}}^{\downarrow}N_{e{_{4}p_{2}}}^{\downarrow}}},
\label{ratios}    
\end{equation}

\noindent where $N_{e{_{i}p_{j}}}^{\sigma}$ represents the number of coincidences in the $i^{th}$ electron detector and the $j^{th}$ proton detector with $\sigma=\uparrow(\downarrow)$ indicating that the neutron polarization is parallel (antiparallel) to the neutron beam.  Each factor can be written as

\begin{equation}
N_{e{_{i}p_{j}}}^{\sigma} 
=C^{\sigma}\Omega^{\sigma}_{e_{i}}\Omega^{\sigma}_{p_{j}}\epsilon^{\sigma}_{e_{i}}\epsilon^{\sigma}_{p_{j}}
\epsilon_{e{_{i}p_{j}}}^{\sigma}(1\pm \vec{K}\cdot \vec{P}D),
\label{eachn}    
\end{equation}

\noindent where $C^{\sigma}$ is proportional to the beam flux for the given neutron polarization direction, the $\Omega$ terms are the solid angles subtended by the indicated detectors, the $\epsilon$ terms are the 
overall efficiencies of the electron and proton detectors, $\epsilon_{e{_{i}p_{j}}}^{\sigma}$ is the correlated coincidence efficiency, and $\vec{P}$ is the average neutron polarization.  The instrumental constant $\vec{K}$ reflects the reduction in sensitivity to the triple correlation due to finite detector solid angles and can be calculated by Monte Carlo simulation for a specific detector geometry.  The ratio is constructed so that the sign on the $\vec{K}\cdot \vec{P}D$ term is ``positive'' for all of the factors in the numerator and ``negative'' for those in the denominator.  The purpose in constructing such a ratio is that the solid angles, individual detector efficiencies, and beam flux factors cancel, separately, for each of the two polarization directions.  Only correlated efficiencies do not cancel exactly.  These will be discussed in the next section. The measured \D-coefficient can be extracted from Equations~\ref{ratios} and~\ref{eachn} to yield

\begin{equation}
D=\frac{1}{\vec{K}\cdot \vec{P}}\frac{(R^{1/8}-1)}{(R^{1/8}+1)}.
\label{finald}    
\end{equation}

\subsubsection{Minimizing False \D-Coefficient Effects}

Correlated efficiencies arise primarily because the spin-neutrino and spin-electron decay coefficients in Equation~\ref{jtweqn} are non-zero.  These decay correlation coefficients couple the neutron spin direction and the momenta of the decay products.  The momenta are in turn coupled to each other via limitations in the available phase-space.  Because correlated efficiencies do not cancel exactly, they can give rise to a nonzero \D. This effect has been observed in a previous experiment~\cite{ERO78}.  There are two important contributions to this type of false asymmetry.  The first can arise from spatial variations in the polarization direction of an extended beam.  This effect limits the maximum size of the beam and places constraints on the uniformity of the polarization direction.  The second effect arises from non-uniformities of the individual detector efficiencies.  Efficiencies can vary either as a function of panel position or the angle at which the decay particles strike the detector faces.  Segmentation of the detector panels helps to reduce these effects by essentially allowing multiple simultaneous experiments.

Known effects contributing to the systematic uncertainty associated with the experiment were evaluated both analytically and with Monte Carlo simulations~\cite{WAS94}.  As a result of these studies strict requirements were placed on the symmetry of the detector system and on the accuracy of the alignment of all beam-line components.  Beam-line components are aligned to better than 2 mrad and mounted kinematically in order to retain the integrity of the alignment when components are removed for access or repairs.  The detector alignment was checked after final beam line buildup and detector placement using a theodolite and temporary cross-hairs.  The magnetic field was aligned prior to detector placement using a fluxgate magnetometer mounted on a translation stage.   The alignment of both the proton and beta detector segments relative to their mounts was set precisely during assembly. Using measured values of $A$ and $B$ and simulating all possible asymmetries, we estimate that the alignment criteria above generates a maximum systematic uncertainty of approximately $10^{-4}$.  This can be confirmed by {\it in situ} systematic tests similar to those performed in the 1997 run~\cite{LJL99}.

\begin{figure*}[t]
\includegraphics[width=5.5in]{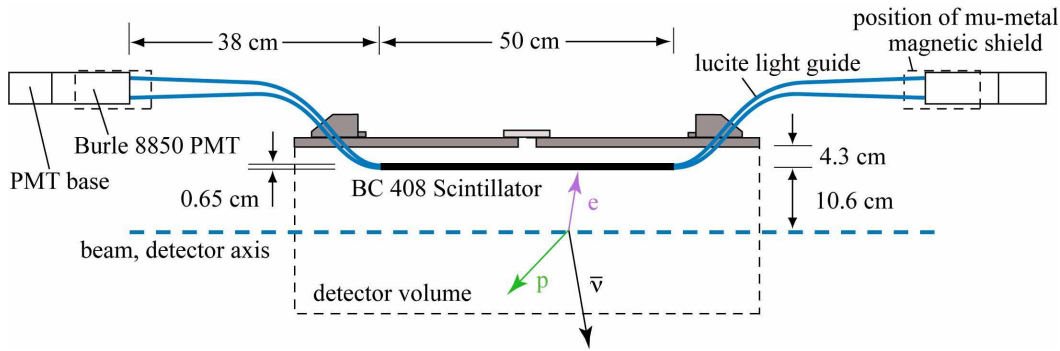}
\caption{Side view of a beta detector.}
\label{betadet}
\end{figure*}
\subsubsection{Statistical Considerations}

The emiT apparatus has a limiting statistical sensitivity of $1/( \vec{K}\cdot \vec{P}\sqrt{N})$, where $N$ is the total number of coincidences, and $\vec{P}$ and $\vec{K}$ are as defined previously.  Because the available beam is limited, it is necessary to maximize the counting rate, which is strongly affected by detector acceptance and configuration.  A larger beam diameter (such as would be obtained with a larger collimator) increases the decay rate by allowing more of the available flux into the detector region.  However, larger beam diameters increase the probability of misidentifying the sign of the triple-correlation and thus decrease the sensitivity per decay; this effect manifests itself quantitatively as a smaller $\vec{K}$ and is the reason behind the beam related reduction in sensitivity discussed previously.  A beam diameter of 6~cm was chosen as an acceptable compromise between these considerations and results in $|K|\approx0.3$.

It is also important to maximize the solid angle of each detector segment.  In the final detector design, the sensitive volume of the array is significantly longer than in previous experiments; moreover, the detector segments are configured with virtually no dead space between them so that both the proton and electron detectors cover nearly $\pi$ radians. 
 
A final important consideration regarding system sensitivity is the rate of background counts.  Subtracting a large background increases the statistical error in the total number of counts and may also impact the counting rate itself by limiting the livetime of the system.  Given the compact configuration of the detectors around the beam and that both the spin flipper and current sheet are activated by neutrons, beam related backgrounds are a serious concern.  These backgrounds are reduced greatly by a compound collimator design and the careful use of lead, concrete, and neutron-absorbing plastic shielding, particularly around the current-sheet spin-flipper.   In addition, the beam stop is located as far from the detector region as possible.  Other sources of noise in the detectors, such as high-voltage related particle emission and thermal noise are addressed by careful electrode polishing and cooling to approximately -100 $^\circ$C respectively.  Hardware energy and software timing cuts further reduce background rates.  The overall beam related background rate in all four beta detectors combined is below 300~s$^{-1}$, while the total rate of background events in the proton detectors is below 2~s$^{-1}$.  Ultimately a signal-to-background of better than one hundred to one is achieved by requiring proton-beta coincidence.

\subsection{Beta Detector Design}\label{betadesign}

The sensitive region of the emiT beta detectors was fabricated from 
slabs of Bicron BC408 plastic scintillator cast to a thickness of 0.64~cm 
and diamond-milled to a rectangular prism measuring 50~cm by 8.4~cm.  
The thickness of 0.64~cm is sufficient to stop a 1~MeV 
electron and is therefore adequate for detecting neutron decay betas, 
which have an endpoint energy of 782~keV. Scintillation photons are transported to either end 
of the scintillator by total internal reflection at the smooth surfaces, where they are ultimately detected by Burle 8850 photomultiplier tubes 
(PMTs).  Although a greater scintillator thickness would have had the 
advantage of presenting lower light-loss (the photons make fewer 
reflections during travel to the PMTs), this apparent advantage is 
outweighed in our case by the reduction in sensitivity to gamma ray 
background presented by a thinner section scintillator.

Each end of the scintillator is adhered to an ultra-pure Lucite light-guide with optical epoxy.  The light-guides adiabatically curve and transform to a circular cross-section, so that they may be optically coupled to the photomultiplier tubes.  The guides curve and pass through an opening in an aluminum housing, which forms a vacuum-tight compression seal against the light-guide with a buna o-ring.  This design permits detector operation with the scintillator positioned inside the detector vacuum chamber while the attached photomultiplier tubes remain external to the vacuum.  Figure~\ref{betadet} presents a side-view of a beta detector assembly.

During the first run of the emiT detector, it was observed that the beta detector exhibited large, continuous rates with occasional large spikes that correlated with breakdown of the proton acceleration electrodes.  Subsequent tests determined that the majority of this rate was from electron field emission at points on the proton detector electrodes.  This effect was mitigated by wrapping the scintillators in a layer of aluminum foil thick enough to stop a 40~keV electron.  In addition, a wrapping of aluminized mylar was placed beneath the foil.  Grounding of both wrappings prevented the detectors from becoming charged.  

PMTs do not function properly in magnetic fields.  Since the uniformity of the guide field in the detector region is critical, both mu-metal cylinders and active shielding were used on the PMTs to minimize stray magnetic fields.  The active shields surround the mu-metal and are configured as a pair of nested, coaxial solenoids of equal length but differing in diameter.  The current passing through each solenoid (in any pair) flows in opposite directions, and its value was chosen so that the resultant magnetic field at the PMT positions is nearly zero.  Outside the solenoid pair the field drops off rapidly since they are designed to have equal and opposite dipole moments.  Thus the field distortion they create at the detector center is an order of magnitude smaller than the distortion the mu-metal shields would otherwise create.  An air-cooling system was installed around the tubes and bases inside the active magnetic shields to maintain temperature stability.  

\subsection{Proton Detector Design}\label{protdesign}

\begin{figure}[t]
\begin{center}
\includegraphics[width=3.0in]{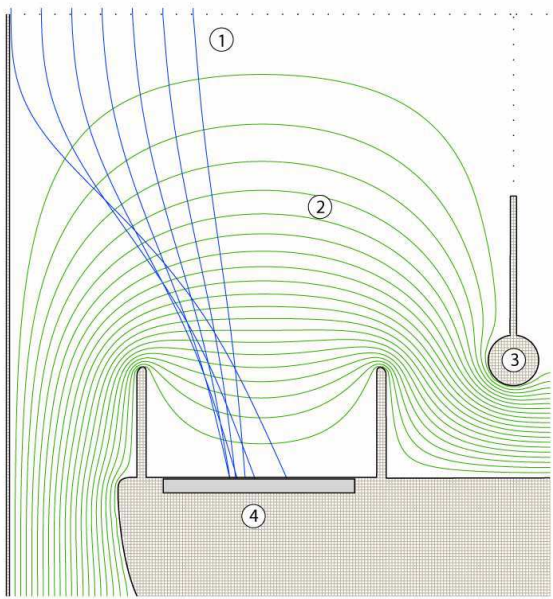}
\end{center}
\caption{Schematic of a proton focusing cell showing (1) sample proton trajectories, (2) equipotential surfaces, (3) ground plane assembly, and (4) location of the surface-barrier detector in the high voltage electrode.  Protons entering the cell at a shallow angle can strike closer to the edge of the detector}
\label{protoncell}
\end{figure}

Each of the four proton segments allowed for the placement of a 2 by 8 array of silicon surface-barrier diode detectors (Ortec~AB-020-300-300-S). The detectors have an active layer 300 mm$^2\times300$~$\mu$m and are positioned behind proton acceleration and focusing cells.  Each cell consists of a grounded box with the top and the upper half of the sides covered by a grounded wire mesh (94\% transmitting) through which the recoil protons enter.  Once inside the box, the protons are accelerated and focused down the center of a cylindrical tube held at a high negative potential that was normally between -25~kV and -32~kV. A schematic representation of a focusing cell is shown in Figure~\ref{protoncell}.  Surface-barrier detectors have a room-temperature leakage current on the order of one microamp and must be cooled to -100~$^\circ$C to achieve the necessary resolution.  Located behind each detector is a preamplifier and cooling attachments.  An optical-fiber link was used to transmit analog signals from the preamps to the data acquisition electronics (see Sections~\ref{fiberlink}~and~\ref{preamp}).

The focusing scheme serves two purposes: accelerating protons (750~eV maximum kinetic energy from neutron beta decay) to the energy needed to penetrate the dead layer and be resolved adequately from the detector noise, and reducing the area of silicon needed to collect the protons.  This reduction of the total diode area by a factor of 7 significantly reduces the noise generated by the detectors as well as the associated cost.  Monte Carlo simulations based on the SIMION~\cite{SIM03} electrostatic modeling code indicate that the recoil protons that enter a cell are focused onto the detectors with very high efficiency (approximately 90\%). Those protons not successfully focused enter very near a cell wall or at shallow incident angles.  A few percent of these scatter, yet still strike the detector and add a small low-energy tail to the proton peak.  The focusing efficiency is independent of the decay position within the beam and is not expected to be a source of systematic errors.  Each proton panel ground plane is segmented longitudinally into eight pairs of square cells 4.0~cm on a side providing the longitudinal segmentation required to further reduce the magnitude of the systematic effects.

An electronics rack is maintained at the same 
high voltage as the proton focusing system.  This rack is isolated by 
a large transformer and housed inside a Faraday cage and protected by 
an interlock system.  Contained in this high-voltage rack are the 
power supplies for the proton detector preamplifiers and fiber-optic 
links as well as the detector bias voltage supply.  The rack also 
contains a protection interlock for the detector bias leakage current 
as well as analog fiber-optic outputs that allow low-voltage 
monitoring of both the total detector leakage current and bias 
voltage.

\subsubsection{Analog Optical Link}\label{fiberlink}

Analog signals from the preamps are transmitted through fiber-optics directly to the Shaper-Analog to Digital Conversion (ADC) cards (see section~\ref{vme}).  This allows the Shaper-ADCs, along with their VME crate and NIM electronics, to be at ground potential with only the power supplies held at high voltage.  These electronics are then completely immune to potential damage resulting from high voltage breakdown.  This approach also reduces the size of the apparatus being floated at high voltage, decreasing its capacitance and the inductance of the lines connecting the high-voltage crate to the proton paddles.   Both tend to reduce the odds of preamp damage.

The analog fiber-optic (F/O) link is built around the Hewlett-Packard HFBR\,1526 Light Emitting Diode (LED) transmitter and HFBR\,2526 receiver.  A fast amplifier (LT1191) is used to drive the LED, and a second one to buffer the receiver into 50 ohms for the Shaper-ADC cards. The transmitter can be driven directly from the existing preamp output, and a receiver was incorporated into the custom Shaper-ADC boards with the option of either F/O or BNC input.

To minimize the contribution of the F/O link to the total noise level, the largest possible gain ahead of the F/O transmitter was desirable consistent with the dynamic range of the receiver and of the expected signals.  See section~\ref{preamp}.  

\subsubsection{Proton Detector Preamplifiers}\label{preamp}

The proton signals generated by the silicon surface-barrier diode detectors are relatively small (approximately 8,300 ion pairs for a proton acceleration potential of 30~kV); moreover, the capacitance of the surface barrier is approximately 80~pF when fully depleted.  Accordingly, it was necessary to design and build low-noise preamplifiers for use with high-capacitance detectors.  The emiT preamp has two gain stages.  The input stage is a folded cascode amplifier with an Interfet IF4501 input Field Effect Transistor (FET), which has a capacitance of 35~pF and g$_{fs}$ of 15~mS.  A 1~pF capacitor in parallel with a 1~G$\Omega$ resistor provide negative feedback.  The second stage consists of a bipolar folded cascode amplifier buffered with a Darlington emitter follower.  The preamp has an overall gain of approximately 9.5~V~pC$^{-1}$ with a 60~ns rise time.  

Due to the potential for damage from high voltage breakdown, care was taken to provide the preamps with some protection against impulse currents.  The output stage was clamped to the -6~V supply and to ground through a Zetex BAV99 dual diode.  In addition, a 50~$\Omega$ resistor was added in series with the output line.  Finally the input FET gate was clamped to the -6~V power supply through an Interfet PAD1 low leakage diode and self clamped to ground through the intrinsic gate-source junction of the FET.

As the leakage current, and hence the performance, of surface-barrier detectors is strongly temperature dependent, considerable attention was given to minimizing the power consumption.  Heat dissipation in vacuum (approximately 15~mW per channel) was aided by the use of high thermal conductivity ceramic substrates (Rogers RO4350B). The overall proton paddle design is a 16 channel motherboard mounted behind the focusing tubes that clamps the detectors in place.  The majority of components are surface mount on interchangeable in-line modules.  

\section{Electronics and Data Acquisition}\label{daq}

\begin{figure}[t]
\includegraphics[width=3.2in]{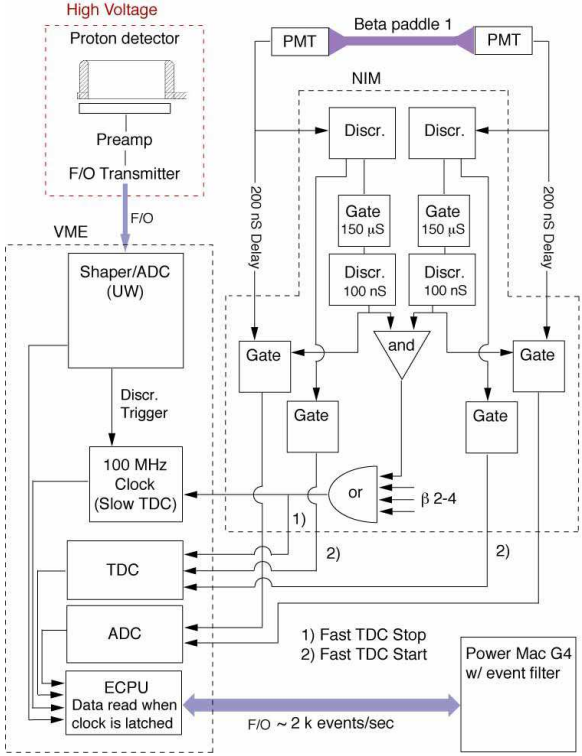}
\caption{A schematic representation of the emiT data acquisition system}
\label{fig:DataFlow}
\end{figure}

The electronics used to run emiT consist of VME and NIM 
electronic modules.   A block diagram of the hardware setup is shown 
in Figure~\ref{fig:DataFlow}.  Analog signals from the proton segment 
preamplifiers are carried by four fiber-optic bundles directly to 
custom VME based shaping and ADC cards.  Two NIM bins contain the 
logic electronics for the beta and proton detectors, spin-flip 
control, and the hardware triggers.  The beta detector PMTs are 
read out by VME based charge integrating ADCs and relative timing of 
the PMTs are recorded in Time to digital Conversion (TDC) units.  A Motorola MVME\,167 embedded 
computer (EPCU) provides real-time readout of the proton detector 
and beta detector electronics, scalers, and also the slowly varying 
analog signals (e.g., magnet currents and applied high voltages) used 
to monitor the status of the experiment.

\subsection{Proton Detector Signal Processing}\label{vme}

The Shaper-ADC boards used to collect data from the surface barrier 
arrays are custom-built eight-channel modules.  Each channel has a 
four-stage integration shaping network and an independently 
controlled level-crossing discriminator.  The peak detection circuit 
consists of a transistor-buffered, dual-differentiating network. 
Shaped pulse conversion is accomplished via a 12-bit ADC and read out 
through the standard VME bus.  Each board also contains scalers that 
can be used to monitor trigger rates on individual channels as well 
as the total board trigger rate.  The boards support an external 
inhibit as well as a fast clear function.  The boards are controlled 
through an Altera 9480 series Field Programmable Gate Array (FPGA) 
and can be run independently or in parallel with additional boards.

The VME boards were designed to accept negative pulses from 5 to 
3000~mV in amplitude depending upon the gain setting.  In order to 
allow adjustment of the dynamic range of the boards, a Maxim MAX\,7524 
multiplying D/A converter is inserted between the first and second 
shaper stages.  This allows for 254 steps of individual channel gain 
adjustment via VME control. Individual channels can also be remotely 
disabled by setting the gain to zero.

The on-board shaping network consists of four successive $0.5~\mu$s 
integrators and a $50~\mu$s  preamplifier tail cancelation circuit. 
The integration operations are done primarily by op-amps.  Peak 
detection is accomplished with a transistor-buffered network.  This 
circuit provides a steep-sloped bi-polar pulse, centered on the peak 
of the shaped signal.  The differentiated signal is amplified using 
an op-amp and fed to a comparator.  The shaped signal is first 
compared to a DC level provided by a programmable buffered Analog Devices AD\,7226 
DAC. A signal of sufficient amplitude then results in the assertion 
of a Sample and Hold (S/H) when the differentiated shaped pulse 
crosses zero.  This method allows for user-defined pulse height 
discrimination with minimal S/H timing walk.

The logic to run the VME interface is programmed directly into the 
Altera FPGA. Only VME line buffers are required outside of the FPGA. 
The interface to VME uses 16 address and 16 data lines.  The board 
design does not include on-board memory and so must be operated in a 
polling readout mode.

\subsection{Beta Detector Signal Processing}\label{vme}The 
individual PMTs charge information as well as the relative timing 
between the two PMTs are recorded for each beta scintillator paddle. 
Charge information is recorded using CAEN Model V862 12 bit 
individually gated QDCs.  The relative timing between the PMTs is 
measured using CAEN Model V775 8 bit TDCs.
Signals from the phototube bases are passively split and one half 
is delayed in cable by 200~ns ($\beta_{\textrm{Delayed}}$).  The 
other half goes through a discriminator whose threshold is set to be 
just above the single photo electron peak.  The output of the 
discriminator is double gated, the first gate of approximately 
150~$\mu$s reduces the effect of after-pulsing, and the second gate 
of 100~ns allows for the transit time of a photon across the length 
of a beta paddle.  This signal is then split, with the first half 
starting a TDC (CAEN Model V775 8 bit TDCs).  The second half is used 
to require a coincidence, $\beta_{\textrm{AND}}$,  between the two 
ends to the paddle (a 100~ns window coming from the previous gate). 
A logic fan-in of the  $\beta_{\textrm{AND}}$s from each paddle is 
used to form a logic or, $\beta_{\textrm{OR}}$. 
$\beta_{\textrm{OR}}$ is used as a trigger for latching a register on 
the latched clock trigger board, as a global stop for the TDCs, and 
as a gate for QDC conversion of $\beta_{\textrm{Delayed}}$.

\subsection{Hardware and Trigger logic} A VME based 
100 MHz latched clock trigger board was designed to provide relative 
timing of both proton and electron singles events.  Two Altera 
programmable logic devices (PLDs) were used.  An EPM9320, 208 pin 20~ns  chip is used to provide the 32 bit VME interface.  An EPF10K10 
series 144 pin 3~ns chip is used for the 56 bit 100 MHz counter and 
five individually latched 56 bit registers.  Four of these registers 
are latched by TTL inputs from either the electron, proton, spin 
flip, or monitoring logic and one register is latched by a software 
command.  Each register has an inhibit output that is asserted when 
the register has been latched.  This inhibit output is used to 
disable further proton or beta signal conversions, until the 
conversion hardware has been read out.  Once the latched register is 
read the inhibit signal returns to logic zero and the conversion of 
the proton or beta signals are once again enabled.  Logic outputs 
reflecting any discriminator fires in the Shaper-ADC boards are combined in a logical OR
to provide a trigger for latching the proton register on the timing 
board.  Data is acquired as follows.  A hardware trigger, for 
example a coincidence between two ends of a beta paddle, causes a 
latch of the beta register of the latched clock trigger card.  This 
sets an inhibit that blocks further triggers in the beta logic.  The 
EPCU polls the timing board on a regular basis and if any register 
has been latched, a read of the appropriate QDCs and TDCs is 
initiated.  The inhibit is then cleared and data acquisition is 
resumed.  A similar process is followed for proton events.  Data is 
written into a local VME dual port memory configured as a circular 
buffer.  The dual port memory, which is located on the SBUS Model 620 
VME to PC controller is simultaneously read out by the data 
acquisition user interface computer, a Macintosh G4.  This computer 
is then displays real time monitoring information and records the 
data to disk. 

\subsection{Monitoring and spin-flip control}A XYCOM 
200 with programmable I/O lines and built in counter timers is 
programmed to generate two periodic timed outputs that initiate 
either a spin-flip sequence or a full monitoring readouts throughout 
the data collection run.  Monitoring of hardware is accomplished 
using VME based 12 bit ADCs (Acromag IP\,220) and scaler counters 
(LeCroy 1151).  Rates in the beta detectors, proton detectors and 
beam fission monitor are tracked in the scalers.  In addition, proton 
paddle temperature, detector bias and leakage current are converted 
to frequency for optical transmission from high voltage, and are also 
tracked in the scalers.  Magnetic coil currents, LN$_2$ fill 
controls, and vacuum status are monitored using the ADC modules. 
Time stamping is accomplished using the latched clock trigger card 
discussed earlier. These data are then output into the event mode data 
stream.  Of the fifty parameters, thirty-five are used for alarms. 
When a value exceeds acceptable bounds a pager based alarm system is 
activated, allowing for a timely maintenance response. 

\section{Beta Detector Performance}\label{betaperform}

The four individual beta detectors were initially tested offline using radioactive sources to determine the optimal operating voltages of their respective pairs of PMTs.  Once operating 
at the appropriate voltage, the detectors were individually tested to obtain characterization parameters for energy calibration, detector light-loss, energy and timing resolution, 
and uniformity of both the hardware trigger and the full-range response.

For each detector, an initial calibration was obtained from the 976~keV conversion electron peak of a $^{207}$Bi source.  Because this peak is normally difficult to resolve in the presence of a large Compton continuum, it was enhanced by placing a thin (0.5~mm.) scintillating disc between the source and the beta detector.  Triggering on pulses from this thin scintillator dramatically reduced the Compton scattered photons seen in the detector and allowed the conversion peak to be well-resolved.  The energy loss in the thin scintillator was measured using a surface-barrier detector.  
Another calibration point came from the 200~keV conversion line of a $^{113}$Sn source, which has greater conversion efficiency, thereby allowing use without an external converter.  After matching PMT gains with the source positioned at the detector's center, each source (in turn) was scanned with two linear translation stages across the face of each detector to measure the response as a function of position along the detector.  The variation of the efficiency as a function of position was shown to be less than a few percent.   In addition, data from this scan was used to measure other parameters such as the light-collection efficiency and attenuation lengths.

\begin{figure}[t]
\begin{center}
\includegraphics[width=3.2in]{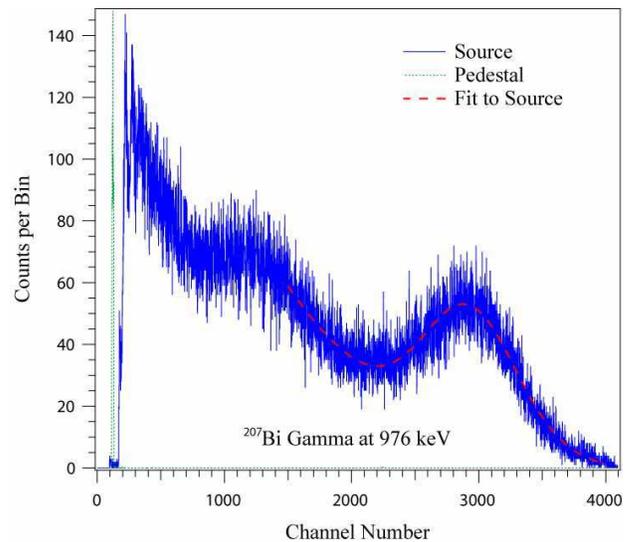}
\end{center}
\caption{A $^{207}$Bi spectrum from a single beta detector taken {\it in situ} without the thin trigger scintillator in place.  The peak is at 976~keV.  The resolution is approximately 18\%.  Also visible in the spectrum is a lower energy set of conversions at 482~keV.  The simple fit is a gaussian with an exponential background}
\label{betaspect}
\end{figure}

A pulsed nitrogen laser system was used to verify the stability of the beta detectors.  Its light output was normalized by using light from a fiber going to a separate scintillation detector with a $^{207}$Bi source as a reference.  During testing, no PMT drift was measured larger than the 5\% accuracy of the monitor.  

Testing the beta detector timing response required sources that would illuminate only a very small region of the scintillator.  For this test we used the laser fiber and a $^{90}$Sr source that was encased in a brass howitzer and released a narrow beam of electrons through a small tunnel on one end.  We measured the full-width at half-maximum (FWHM) of the end-to-end relative time for various energies as a function of discriminator level and saw that a minimum was reached at 
all energies for thresholds of 100 mV and below.  For higher energies near the neutron beta endpoint, this minimum in the width was approximately 0.5~ns and increased to about 1.3~ns at 200~keV. Measuring the shift of the time peak as we moved the source gave an effective velocity, $\nu_{eff}$, for the scintillation light of about 16~cm/ns, which allows us to extract beta position with a resolution between 4~cm and 10~cm, given by $\Delta x = 2\nu_{eff}\Delta t$. 


The energy resolution of the segments was measured using the conversion sources.  Immediately after fabrication, a FWHM of 15\% was observed at about 1~MeV.  The PMT gains were set to place the 976~keV conversion electron peak of $^{207}$Bi at about channel 3000 of the 4096 channel 12 bit ADCs.  {\it In situ} measurements show a resolution of about 18\% at this energy.  A random trigger was used to determine the position of the pedestal, giving a second calibration point.  Using this calibration, the beta energy thresholds were determined to be about 35 keV for three of the paddles and 50~keV for the fourth.  The fourth paddle requires a higher threshold because the dark current in one of the phototubes is higher than expected.

When installed on the beam line, each beta detector has a trigger rate of about 40~s$^{-1}$ when the local beam shutter is closed and the proton detector high voltage set to zero.  This noise is due primarily to reactor related backgrounds that penetrate through the shielding, but also comes from cosmic rays and dark current in the photomultiplier tubes.  Opening the local beam shutter increases the total singles rates to about 300~s$^{-1}$ including decay electrons.  In contrast to the first run of emiT, the background is not dependent on the proton detector acceleration voltage.

\section{Proton Detector Performance}\label{protperform}
\begin{figure*}[t]
\begin{center}
		\hspace{-0.25in}
	\raisebox{2.4in}{(a)\ }
	\includegraphics[width= 2.5in]{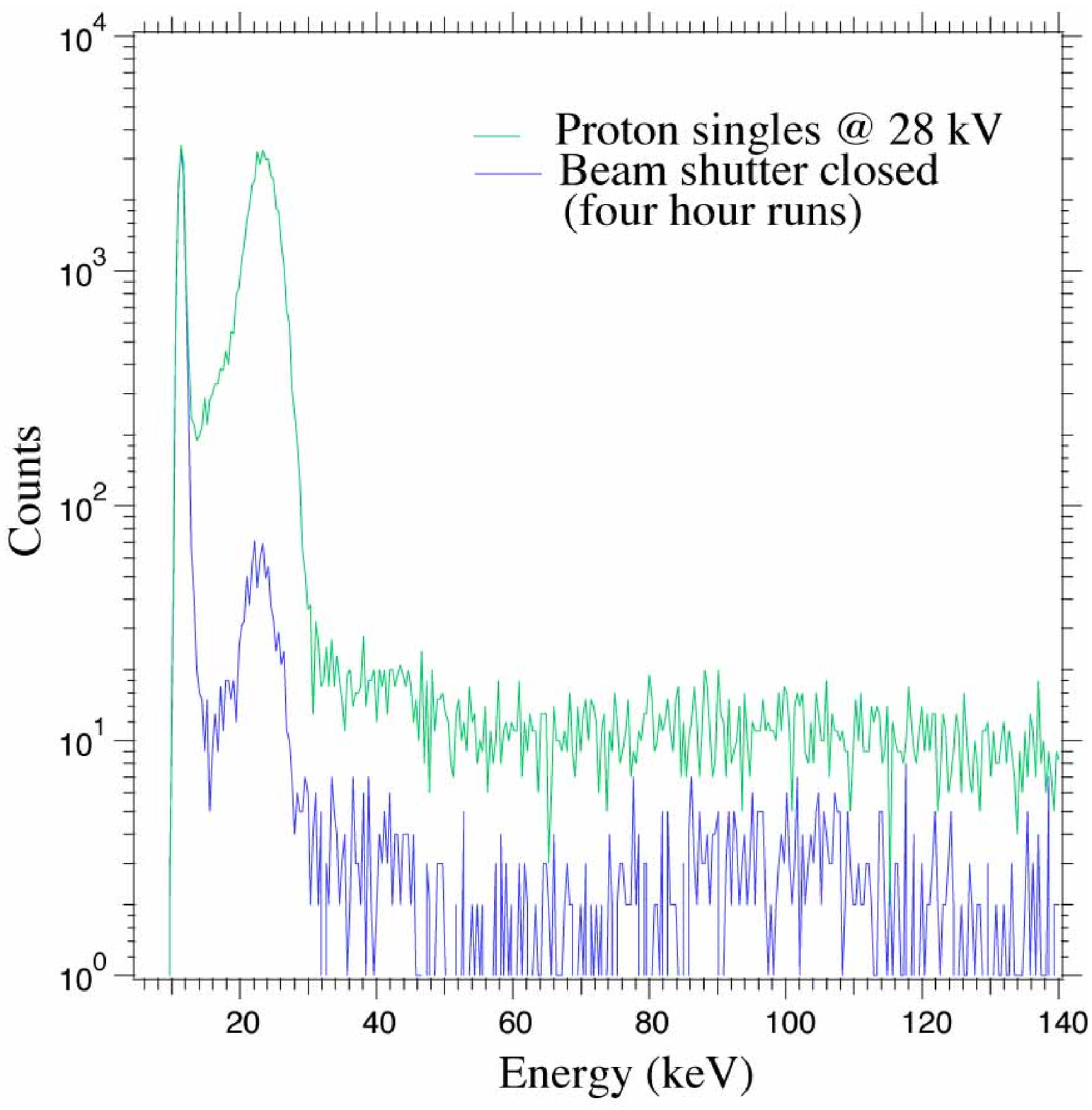}
	\hspace{.1in}
	\raisebox{2.4in}{(b)\ }
	\includegraphics[width= 2.5in]{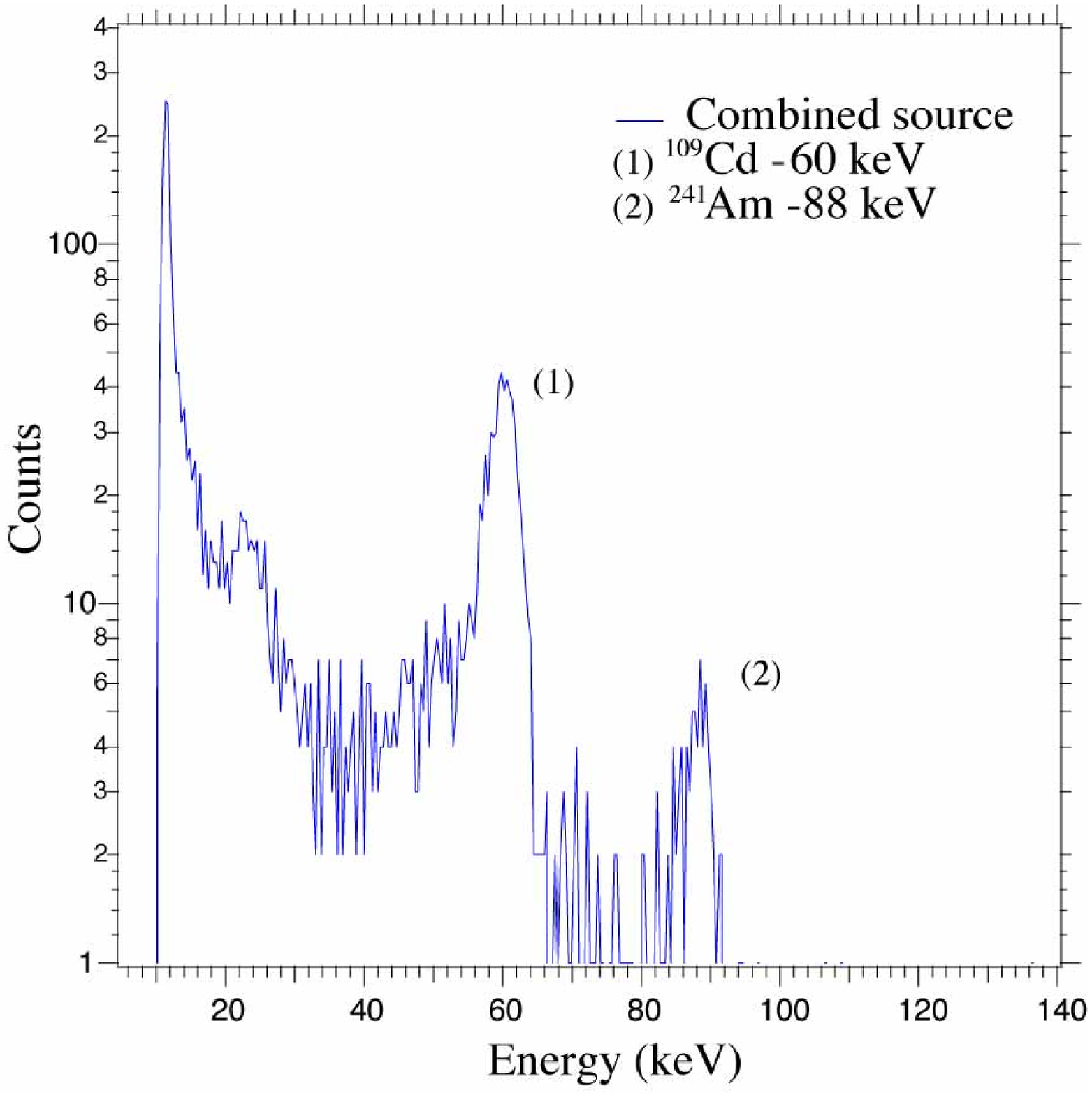}
	\vspace{.1in}
	\par
	\hspace{-0.25in}
	\vspace{.1in}
	\raisebox{2.4in}{(c)\ }
	\includegraphics[width= 2.5in]{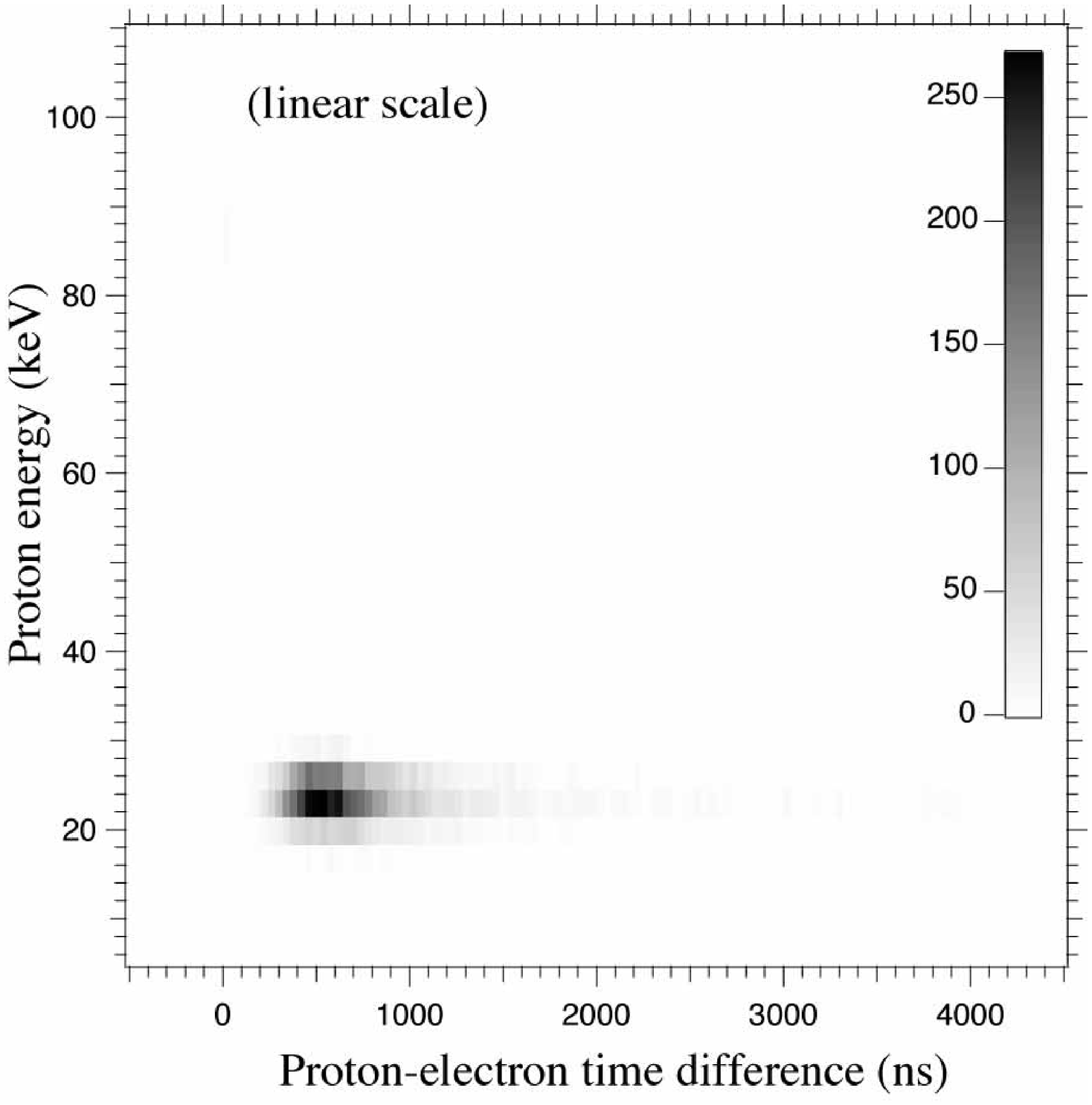}
	\hspace{.1in}
	\raisebox{2.4in}{(d)\ }
	\includegraphics[width= 2.5in]{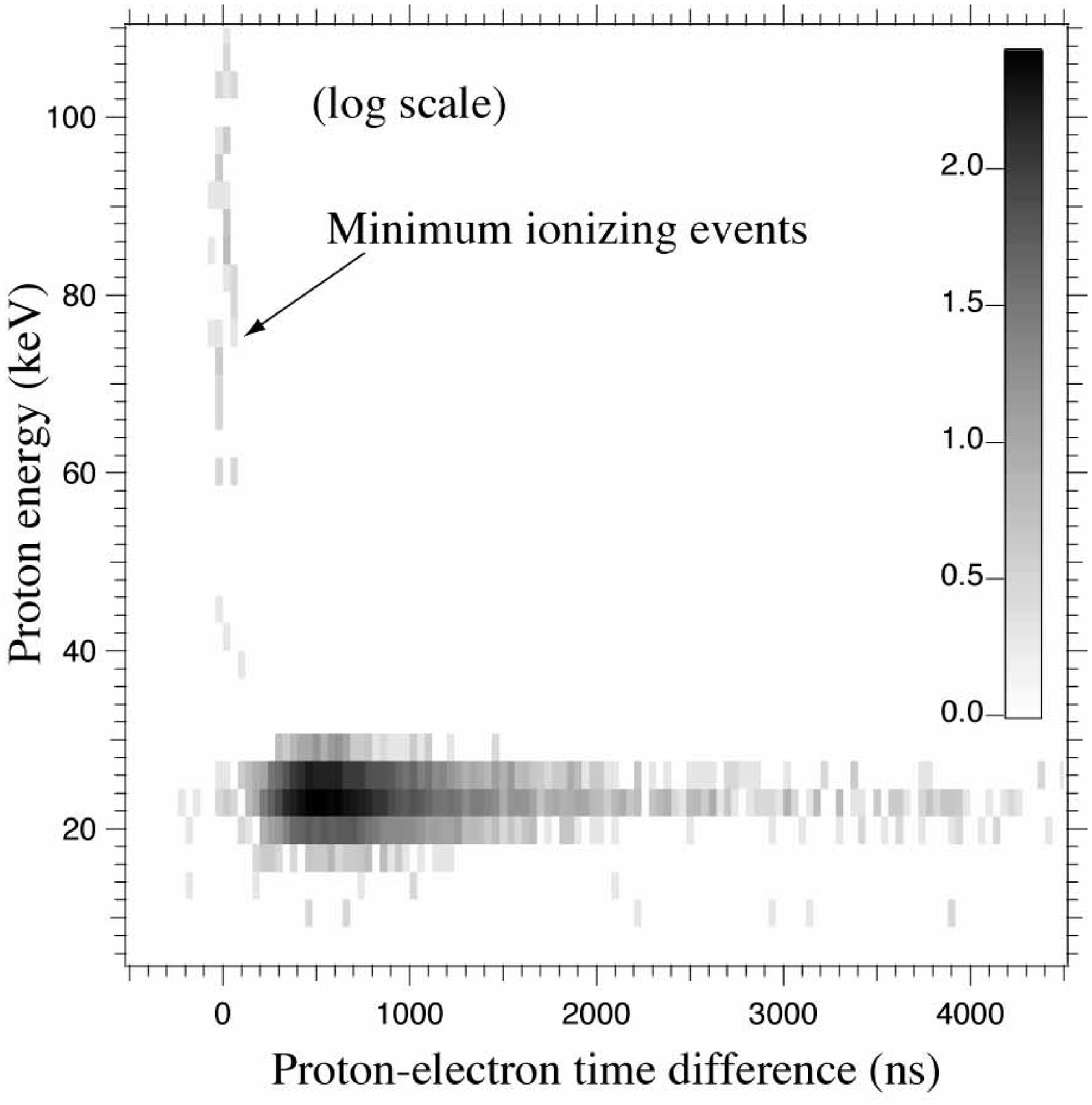}
	\par
	\caption{Proton energy calibration data from one of the highest resolution surface-barrier detectors.  (a) Spectrum from $^{241}$Am and $^{109}$Cd sources. The FWHM is 4.78 keV at the 59.5 keV line.   (b) Pulse height singles spectrum of 28 kV decay protons for a typical four hour run.  The large peak in the beam-off data is from high voltage protons and is below 0.6 s$^{-1}$ in all channels.  The lower amplitude peak at approximately 100~keV is the minimum ionizing peak in silicon. (c) Coincidence plot of the same data, proton energy vs delay time. (d) Log plot of the same data showing prompt from minimum ionizing particles and very low backgrounds.}
\label{fig:calibrations}
\end{center}
\end{figure*}

The surface-barrier detectors were specially manufactured to have an entrance window of 20~$\mu$g Au, half the normal thickness.  This is important because it reduces energy loss in the insensitive region (dead layer) of the detector.  The reduced energy loss allows lower acceleration voltages, decreasing the rate of high-voltage breakdown and reducing high-voltage associated backgrounds.  Surface-barrier detectors were chosen to replace the PIN diodes used in the first run because low energy protons lose a calculated 2~keV  in the 20~$\mu$g gold front entrance window (electrical contact), compared to 12 keV in the same thickness of Si.  Clearly this also helps to mitigate potential high-voltage related problems by also allowing operation at lower acceleration voltages.  To verify the performance and dead-layer thickness of the proton detectors prior to moving the apparatus to the neutron beam line, we constructed a simple duo-plasmatron source to produce a low-energy (0-few hundred eV) proton beam.  The source attached to the downstream end of the detector.  The proton beam was collimated and entered the chamber where it Rutherford scattered off a movable Al target and struck the detectors~\cite{Naab02}.  

In addition to careful characterization of the surface-barrier detectors, an extensive investigation of the unexpected high-voltage related background seen during the first run was carried out.  It was found that high electric fields around the edge of the focusing tubes caused electron field emission.  This results in two sources of background: bremsstrahlung and ionization of adsorbed hydrogen on the focusing assembly ground plane.  These ions are accelerated back into the detectors in exactly the same manner as a decay proton, yielding a background that cannot be removed by cuts on proton energy.  To eliminate these events, the focusing tubes were redesigned to produce lower electric fields and carefully polished.  As a result, this source of background has been nearly eliminated (see Figure \ref{fig:calibrations}(a)).  The rate of these background protons varies somewhat from channel to channel, however during typical running conditions it is never above 0.06 s$^{-1}$ per channel.

During testing and early running a large number of surface-barrier detectors suffered from a variety of unexplained problems including abnormally high leakage current and breakdown at voltages well below nominal operating voltage.  However, after biased operation in vacuum for times on the order of a month most detectors behave well.  Energy resolution ranges from 4.5~keV to 8~keV and is sufficient for our purposes.  Typical energy loss is about 5~keV.  We expect this to be larger than the calculated value because of additional losses in inactive silicon directly below the gold entrance window.

 Figure~\ref{fig:calibrations}(b) shows an energy calibration spectrum using an $^{241}$Am source together with a $^{109}$Cd source.  The FWHM of the 59.5 keV line is 4.78 keV. Figure~\ref{fig:calibrations}(a) gives an example of a 28 kV proton energy spectrum, taken without the requirement of an electron coincidence.  The peak  is made up of protons originating almost entirely from neutron decay.  The singles rate in each proton detectors is approximately 3 s$^{-1}$. 
 
In Figures~\ref{fig:calibrations}(b) and \ref{fig:calibrations}(d) shows proton events correlated with a beta trigger.  In  Figure~\ref{fig:calibrations}(b) (logarithmic scale) one can see a prompt background signal (vertical stripe).  This is associated with the beam and the peak energy and spectral shape is consistent with minimum ionizing particles in 300 $\mu$m of silicon.  The energy of the peak in the prompt spectrum is about 100 keV and is very well separated from the decay protons.  This peak is also visible in Figure~\ref{fig:calibrations}(b) as singles events.  At the same energy as the proton peak, but not clearly resolved in the plot due to its small amplitude, is a band of recoil protons for which the coincident electron was not detected.  This band makes up the dominant background.  The coincidence efficiency varies across the front face of the proton paddle but was calculated to be approximately 20\%.  Coincidence events are seen to be in the range of 0.55 s$^{-1}$ to 0.62 s$^{-1}$ for the paddle end and center respectively.   The overall coincidence rate is greater than 30~s$^{-1}$ with a signal to background of better than 100 to 1.

It is clear that the performance of the detector has been greatly improved since the first run.  As shown above, the data quality is high and all known  systematic effects are under control.  We expect that the current run will reach a statistically limited sensitivity to \D\ of $2 \times 10^{-4}$.  Systematic uncertainties should be below $1\times10^{-4}$.

\section*{Acknowledgments}\label{acks}

We acknowledge the support of the National Institute of Standards and Technology, U.S. Department of Commerce, in providing the neutron facilities used in this work.  Thanks are due to Allan Myers, Tim Van Wechel, Doug Will and John Amsbaugh
for technical support and to Mark Wehrenberg for his work on the project.  This research was made possible in part
by grants from the U.S. Department of Energy and the National Science Foundation.  C.~A.~Trull enjoys support from the Louisiana Board of Regents, BoRSF, under agreement NASA/LEQSF(2001-2005)-LaSPACE and NASA/LaSPACE.


\begin{references}
\bibitem{FLIP}The measured flipping ratio was higher in the first run ($R = 11.84\pm0.08$).  This is possibly due to both the cold source upgrade and a shortened Be guide. These changes shift the spectrum to higher wavelength which lowers the polarizing power of the supermirror.
\bibitem{Disclaimer} Certain trade names and company products are mentioned in
the text or identified in illustrations in order to adequately
specify the experimental procedure and equipment used. In no
case does such identification imply recommendation or endorsement
by National Institute of Standards and Technology,
nor does it imply that the products are necessarily the best
available for the purpose.
\end{references}
\end{document}